\documentclass[12pt,twoside]{article}
\usepackage{epic}
\usepackage{eepic}
\usepackage[mathscr]{eucal}
\usepackage{amsmath,amsfonts,amssymb,amsthm}
\usepackage[matrix,arrow]{xy}
\usepackage{times}
\usepackage{pdfsync}
\usepackage{graphicx}% Include figure files
\usepackage{epstopdf}
\usepackage{multibox}
\usepackage{dcolumn}
\usepackage{hyperref}

\def\d_Vphi{\text{d}_V\hspace{-0.06em}\phi}
\def\d_Vphibar{\text{d}_V\hspace{-0.06em}\bar\phi}
\def\d_Vxi{\text{d}_V\hspace{-0.06em}\xi}

\def\be{\begin{eqnarray}}
\def\ee{\end{eqnarray}}
\def\beann{\begin{eqnarray*}}
\def\eeann{\end{eqnarray*}}
\def\beq{\begin{equation}}
\def\eeq{\end{equation}}
\def\ba{\begin{array}}
\def\ea{\end{array}}
\def\ben{\begin{enumerate}}
\def\een{\end{enumerate}}
\def\bea{\begin{eqnarray}}
\def\eea{\end{eqnarray}}

\def\5{\bar }
\def\6{\partial }
\def\7{\hat }
\def\4{\tilde }
\advance\textheight\topskip
\renewcommand{\tilde}{\widetilde}
\renewcommand{\hat}{\widehat}

\renewcommand{\simeq}{\cong}

\newcommand{\dd}{\partial}
\renewcommand{\d}{\partial}

\newcommand{\binner}[2]{%
  {\langle}\kern-4.15pt{\langle}#1{,}\,#2{\rangle}\kern-4.15pt{\rangle}}

\newcommand{\half}{\frac{1}{2}}

\newcommand{\ffrac}[2]{\raisebox{.5pt}%
  {\footnotesize$\displaystyle\frac{#1}{#2}$}\kern1pt}

\newcommand{\dover}[2]{\ffrac{\dd #1}{\dd #2}}

\def\cF{\mathcal{F}}

\def\cN{\mathcal{N}}

\DeclareFontFamily{OT1}{rsfs}{} \DeclareFontShape{OT1}{rsfs}{m}{n}{
<-7> rsfs5 <7-10> rsfs7 <10-> rsfs10}{}
\DeclareMathAlphabet{\mycal}{OT1}{rsfs}{m}{n}
\def\scri{{\mycal I}}%
\begin{document}

\def\mytitle{Supertranslations call for superrotations}

\pagestyle{myheadings} \markboth{\textsc{\small Barnich, Troessaert}}{%
  \textsc{\small Supertranslations and superrotations}} \addtolength{\headsep}{4pt}

\begin{flushright}\small
ULB-TH/11-02\end{flushright}

\begin{centering}

  \vspace{1cm}

  \textbf{\Large{\mytitle}}

%\vspace{1cm}

%{\huge Notes}

  \vspace{1.5cm}

  {\large Glenn Barnich$^{a}$ and C\'edric Troessaert$^{b}$}

\vspace{.5cm}

\begin{minipage}{.9\textwidth}\small \it \begin{center}
   Physique Th\'eorique et Math\'ematique\\ Universit\'e Libre de
   Bruxelles\\ and \\ International Solvay Institutes \\ Campus
   Plaine C.P. 231, B-1050 Bruxelles, Belgium \end{center}
\end{minipage}

\end{centering}

\vspace{1cm}

\begin{center}
  \begin{minipage}{.9\textwidth}
    \textsc{Abstract}. We review recent results on symmetries of
        asymptotically flat spacetimes at null infinity. In higher
        dimensions, the symmetry algebra realizes the Poincar\'e
        algebra. In three and four dimensions, besides the
        infinitesimal supertranslations that have been known since the
        sixties, the algebras are evenly balanced because there are
        also infinitesimal superrotations. We provide the
        classification of central extensions of $\mathfrak{bms}_3$ and
        $\mathfrak{bms}_4$. Applications and consequences as well as
        directions for future work are briefly indicated.
  \end{minipage}
\end{center}

\vspace{1cm}

{\em Proceedings of the {\bf Workshop on Non Commutative Field Theory and
  Gravity}, September 8-12, 2010, Corfu Summer Institute on Elementary
Particles and Physics, Greece and the {\bf 7th Spring School and
  Workshop on Quantum Field Theory and Hamiltonian Systems}, 10-15 May
2010, Craiova and Calimanesti, Romania.}

\vfill

\noindent
\mbox{}
\raisebox{-3\baselineskip}{%
  \parbox{\textwidth}{\mbox{}\hrulefill\\[-4pt]}}
{\scriptsize$^a$Research Director of the Fund for Scientific
  Research-FNRS. E-mail: gbarnich@ulb.ac.be\\$^b$ Research Fellow of
  the Fund for Scientific Research-FNRS. E-mail: ctroessa@ulb.ac.be}

\thispagestyle{empty}
\newpage

\section{The $\mathfrak{bms}_n$ algebra in higher dimensions} 

When studying asymptotic symmetries, a fast way to get an idea of what
the eventual algebra might be is to solve the Killing equation for the
background metric to leading order. When this is done for
asymptotically flat space-times at null infinity in $n$ spacetime
dimensions \cite{Barnich:2006avcorr}, one finds that the symmetry
algebra consists of the semi-direct sum of conformal Killing vectors
of the $n-2$ sphere acting on the ideal of infinitesimal
supertranslations, which are parametrised by arbitrary functions on
the $n-2$ sphere. 

If $x^A$, $A=2,\dots, n$ are coordinates on the $n-2$ sphere, $\bar
D_A $ the associated covariant derivative, $Y^A(x^B)\d_A$ the
conformal Killing vectors and $T(x^A)$ the functions parametrising
the infinitesimal supertranslations, the $\mathfrak{bms}_n$ algebra is
explicitly defined through the commutation relations
$[(Y_1,T_1),(Y_2,T_2)]=(\hat Y,\hat T)$ where
\begin{equation}
  \left\{\begin{array}{l}
      \label{eq:5a}\hat Y^A= Y^B_1\d_B
Y^A_2-Y^B_2\d_B Y^A_1,\\
\hat T=Y^A_1\d_A
  T_2-Y^A_2\d_A T_1 +\frac{1}{n-2} (T_1\bar D_AY^A_2-T_2\bar
  D_A Y^A_1)\,.
\end{array}\right.
\end{equation}

For $n > 4$, the first factor is isomorphic to the $n(n-1)/2$
dimensional algebra $\mathfrak{so}(n-1,1)$ of infinitesimal conformal
transformations of Euclidean space in $n-2$ dimensions and also to
the Lorentz algebra in $n$ dimensions.

When making a more detailed analysis taking the precise definitions of
asymptotically flat spacetimes in higher dimensions into account, it
turns out that the supertranslations collapse to ordinary translations 
so that the resulting symmetry algebra is just the Poincar\'e
algebra
\cite{Hollands:2003ie,Tanabe:2009va}. 

In the realizations of asymptotic symmetries in general relativity in 
higher dimensions, there thus remain only standard rotations,
including the hyperbolic ones, and translations. In three and four
dimensions however, the asymptotic symmetry algebras are
infinite-dimensional and thus yield much more information on the
system.

Before turning to the algebra in three and four dimensions, we make a
couple of remarks on how to actually compute the asymptotic symmetry
algebra and on its realizations.

\section{Asymptotic versus complete gauge fixations. Realizations}
\label{sec:asympt-vers-compl}

There are basically two attitudes to the problem. On the one hand, one
can fix the coordinate freedom only asymptotically in which case the
asymptotic symmetry algebra appears as the quotient algebra of allowed,
modulo an ideal of trivial, infinitesimal transformations. The
advantage of this approach is that it is easier to show that specific
solutions to the equations of motion are admissible, i.e.,
asymptotically flat in the case of interest here. When one chooses to
fix the coordinate freedom completely on the other hand, the
asymptotic symmetry algebra appears as the residual ``global''
symmetry algebra after gauge fixing and no longer depends on
arbitrary functions of the bulk spacetime. The advantage of this
``reduced phase space'' approach is that only physical degrees of
freedom remain. A standard example illustrating this difference is the
Brown-Henneaux \cite{Brown:1986nw} versus the Feffermann-Graham
\cite{fefferman:1985,graham:1991} definition of asymptotically anti-de
Sitter spacetimes in three dimensions.

Recently in \cite{Barnich:2009se,Barnich:2010eb}, we have followed the
latter approach in the asymptotically flat case by using a
Bondi-Metzner-Sachs type of gauge in four dimensions
\cite{Bondi:1962px,Sachs:1962wk,Sachs2} and a reasonable analog
thereof in three. In particular, the asymptotic symmetry algebras to
be discussed below have explicitly been shown to be the same whether
one fixes the gauge completely or only asymptotically. Furthermore, as
suggested by Penrose's conformal approach to asymptotically flat
spacetimes \cite{PhysRevLett.10.66,penrose:1964,Penrose:1974}, we have
considered classes of gauge fixations differing by a choice of the
conformal factor for the degenerate metric on Scri and have
investigated the behavior of the theory under changes of such gauges.

From this point of view, the Newman-Unti (NU) approach to
asymptotically flat spacetimes \cite{newman:891} corresponds to a
different gauge choice for the radial coordinate. In view of its
embedding in the widely used Newman-Penrose formalism
\cite{newman:566} and its direct relevance in many applications, see
e.g.~the review article \cite{newman:1980xx}, it is worthwhile to show
that the asymptotic symmetry algebra is unchanged and to provide
explicit formulae for the realization of the algebra in this
gauge. This has been done in \cite{barnich:2011ty}.

A novel result in our study concerns the realization of the asymptotic
algebra not only on the boundary Scri but in the bulk gauge fixed
spacetime by using a natural modification of the Lie bracket for
vector fields that depend on the metric and is related to the theory
of Lie algebroids \cite{Barnich:2010xq}. Furthermore, this modified
bracket is also needed for the realization on Scri in order to
disentangle the gauge transformations from the residual global
symmetries when allowing for changes of the conformal factor. We have
also studied in detail how the symmetry algebra is realized on the
arbitrary functions parametrizing solution space.

\section{The $\mathfrak{bms}_3$ algebra}
\label{sec:three-dimens-case}

The $\mathfrak{bms}_3$ algebra consists of the algebra of vector
fields on the circle acting on the functions of the circle and has
been originally derived in the context of a symmetry reduction of
four dimensional gravitational waves
\cite{Ashtekar:1996cm,Ashtekar:1996cd}.

More precisely, let $y=Y\dover{}{\phi} \in {\rm Vect}(S^1)$
be the vector fields on the circle and $T(d\phi)^{-\lambda}\in \cF_{\lambda}(S^1)$ tensor
densities of degree $\lambda$, which form a  module of the Lie
algebra ${\rm Vect} (S^1)$ for the  action 
\begin{equation}
\rho(y) t=
(YT^\prime-\lambda Y^\prime T)d\phi^{-\lambda}\label{eq:8}\,.
\end{equation}
The algebra $\mathfrak{bms}_3$ is the semi-direct sum of ${\rm Vect}
(S^1)$ with the abelian ideal $\cF_{1}(S^1)$, the bracket between
elements of ${\rm Vect}(S^1)$ and elements
$t=Td\phi^{-1}\in\cF_{1}(S^1)$ being induced by the module action,
$[y,t]=\rho(y)t$.

Consider the associated complexified Lie algebra and let
$z=e^{i\phi}$, $m,n,k ...\in \mathbb Z$. Expanding into modes,  
$y=a^n l_n $, $t=b^n t_n$,  where
\[
l_n=e^{in\phi}\dover{}{\phi}=iz^{n+1}\dover{}{z}
,\quad t_n= e^{in\phi}(d\phi)^{-1}=i z^{n+1}(dz)^{-1}\,,\] 
the commutation relations read explicitly
\begin{equation}
i[l_m,l_n]=(m-n) l_{m+n},\quad i[l_m, t_n]=(m-n) t_{m+n},\quad i[t_m, t_n]=0\,.
\label{eq:3}
\end{equation}
The non-vanishing structure constants of $\mathfrak{bms}_3$ are thus
entirely determined by the structure constants $[l_m,l_n]=-i f^k_{mn}
l_k$, $f^k_{mn}=\delta^k_{m+n}(m-n)$ of the Witt subalgebra
$\mathfrak w$ defined by the linear span of the $l_n$.

Up to equivalence, the most general central extension of
$\mathfrak{bms}_3$ is given by
\begin{eqnarray} 
\left\{\begin{array}{l}
i[l_m,l_n]=(m-n)
  l_{m+n}+\frac{c_1}{12}m(m+1)(m-1)\delta^0_{m+n}, \cr i[l_m,
t_n]=(m-n) t_{m+n}+\frac{c_2}{12}m(m+1)(m-1)\delta^0_{m+n},\cr i [t_m,
t_n]=0\,.
\label{eq:3a}
\end{array}\right.
\end{eqnarray}
The proof follows by generalizing the one for the Witt algebra
$\mathfrak w$, which is textbook material, see
e.g~\cite{Fuks:1986,Henneauxstrings,deAzcarraga:1989dm}.
Nevertheless, in order to be self-contained, we give a complete
derivation in the appendix.

The associated classical charge algebra of asymptotically flat three
dimensional space-times has been constructed in
\cite{Barnich:2006avcorr} with central charges\footnote{Note that in
  the equivalent gauge fixed derivation given in
  \cite{Barnich:2010eb}, there is a misprint in the last line of
  (3.18), where $\Theta$ has to be replaced by $\Theta+1$.}
\begin{equation}
c_1=0,\quad c_2=\frac{3}{G}\label{eq:7}\,.
\end{equation}

When one considers the extension of the $\mathfrak{bms}_3$ algebra
obtained by replacing the vanishing commutators of the $t_m$'s in
(\ref{eq:3a}) through
\begin{equation}
  \label{eq:1}
  i[t_m,t_n]=\frac{1}{l^2} (m-n) l_{m+n}\,,
\end{equation}
and defines $l^{\pm}_m=\half( l t_{\pm m}\pm l_{\pm m})$, the resulting algebra
turns into two copies of the Virasoro algebra with central charges
$c^\pm=\frac{3l}{2G}$,
\begin{equation}
  \label{eq:2}
 i [l^\pm_m,l^\pm_n]=(m-n)l^\pm_{m+n}+\frac{c^\pm}{12}m(m+1)(m-1)\delta^0_{m+n},\quad
 i [l^\pm_m,l^\mp_n]=0\,,
\end{equation}
which is precisely the value of the classical central extensions in
the charge algebra of asymptotically anti-de Sitter spacetimes
\cite{Brown:1986nw}. In other words, starting from the charge algebra
(\ref{eq:2}) in asymptotically anti-de Sitter space-times, the flat
result is obtained by first writing the algebra in terms of the new
generators $l_m=l_m^+-l_{-m}^-$, $t_m=\frac{1}{l}(l_m^++l_{-m}^-)$ and
then taking $l\to \infty$.

An important question is a complete understanding of the physically
relevant representations of $\mathfrak{bms}_3$. Note that in the
present gravitational context, the Hamiltonian is associated with
$t_0$, so that one is especially interested in representations with a
lowest eigenvalue of $t_0$. This question should be tractable, given
all that is known on both the Poincar\'e and Virasoro subalgebras of
$\mathfrak{bms}_3$.

It turns out that $\mathfrak{bms}_3$ is isomorphic to the Galilean
conformal algebra in $2$ dimensions $\mathfrak{gca}_2$
\cite{PhysRevLett.105.171601}. In a different context, a class of
non-unitary representations of $\mathfrak{gca}_2$ have been studied in
some details \cite{Bagchi:2009pe}.

\section{The $\mathfrak{bms}_4$ algebra}
\label{sec:four-dimens-case}

In four dimensions, the infinitesimal Lorentz transformations appear
as the conformal Killing vectors of the $2$ sphere. By the standard
argument, when focusing on infinitesimal local transformations that
are not required to be everywhere regular, the conformal Killing
vectors are given by two copies of the Witt algebra, so that besides
supertranslations, there now also are superrotations.

More precisely, in stereographic coordinates
$\zeta=e^{i\phi}\cot{\frac{\theta}{2}} $ and $\bar \zeta$ for the $2$
sphere with $\varphi_0=\ln{\half(1+\zeta\bar\zeta)}$, the algebra may
be realized through the vector fields $y=Y(\zeta)\d $, $\bar y=\bar
Y(\bar \zeta)\bar \d$ where $\d= \dover{}{\zeta}$, $\bar
\d=\dover{}{\bar \zeta}$.  If $T(\zeta,\bar\zeta)=\tilde
T(\zeta,\bar\zeta)e^{-\varphi_0}$, they act on tensor densities
$\cF_{\half,\half}$ of degree $(\half,\half)$,
\begin{equation}
t=\tilde T(\zeta,\bar\zeta)e^{-\varphi_0}(d\zeta)^{-\half}(d\bar\zeta)^{-\half}\,,\label{eq:11}
\end{equation}
through
\begin{eqnarray}
\rho(y) t &=& (Y\d \tilde T-\half \d Y
\tilde T)e^{-\varphi_0}(d\zeta)^{-\half}(d\bar\zeta)^{-\half}\,,\\
\rho(\bar y) t&=&(\bar
Y\bar \d \tilde T-\half \bar \d \bar Y
\tilde T)e^{-\varphi_0}(d\zeta)^{-\half}(d\bar\zeta)^{-\half}\,. \label{eq:9}
\end{eqnarray}
The algebra $\mathfrak{bms}_4$ is then the semi-direct sum of the
algebra of vector fields $y,\bar y$ with the abelian ideal
$\cF_{\half,\half}$, the bracket being induced by the module action,
$[y,t]=\rho(y)t$, $[\bar y,t]=\rho(\bar y) t$.

When expanding $y=a^nl_n$, $\bar y=\bar a^n\bar l_n$, $t=b^{m,n} T_{m,n}$, 
with
\begin{equation}
  \label{eq:10}
  l_n=-\zeta^{n+1}\d,\quad \bar l_n=-\bar \zeta^{n+1}\bar\d,\quad 
T_{m,n}=\zeta^m\bar\zeta^n
e^{-\varphi_0}(d\zeta)^{-\half}(d\bar\zeta)^{-\half}\,,
\end{equation}
the enhanced symmetry algebra reads
\begin{eqnarray}
  \label{eq:37}
&  [l_m,l_n]=(m-n)l_{m+n},\quad [\bar l_m,\bar l_n]=(m-n)\bar
  l_{m+n},\quad [l_m,\bar l_n]=0, \cr
& [l_l,T_{m,n}]=(\frac{l+1}{2}-m)T_{m+l,n},
\ [\bar l_l,T_{m,n}]= (\frac{l+1}{2}-n)T_{m,n+l},\ [T_{m,n},T_{o,p}]=0,
\end{eqnarray}
where $m,n\dots\in \mathbb Z$. The Poincar\'e algebra is the
subalgebra spanned by the generators $T_{0,0}$, $T_{0,1}$, $T_{1,0}$,
$T_{1,1}$ for ordinary translations and $l_{-1},l_0,l_1,\bar
l_{-1},\bar l_{0},\bar l_1$ for ordinary (Lorentz) rotations.

The quotient algebra of $\mathfrak{bms}_4$ by the abelian ideal of
infinitesimal supertranslations is no longer given by the Lorentz
algebra but by two copies of the Witt algebra. It follows that the
problem with angular momentum in general relativity
\cite{winicour:1980aa}, at least in its group theoretical formulation,
disappears as now the choice of an infinite number of conditions is
needed to fix an infinite number of rotations. For a complete
analysis, the associated charges and their algebra is needed. This
will be discussed in detail elsewhere. 

In the appendix, we will show that the only non trivial central
extensions of $\mathfrak{bms}_4$ are the usual central extensions of
the 2 copies of the Witt algebra, i.e., they appear in the commutators
$[l_m,l_{-m}]$ and $[\bar l_m,\bar l_{-m}]$.  Contrary to three
dimensions, there are no central extensions involving the generators
for supertranslations.

\section{Outlook}
\label{sec:outlook}

The most obvious questions to be addressed next are a complete study
of the physically relevant representations of $\mathfrak{bms}_3$ and
of $\mathfrak{bms}_4$ and the construction of the surface charge
algebra associated with supertranslations and superrotations in $4$
dimensions. We have recently made progress on the latter problem and
will report on these results elsewhere. It turns out that the
extension in the charge algebra depends explicitly on the fields
characterizing solution space, it is a Lie algebroid $2$-cocycle
rather than a Lie algebra $2$ cocycle.

The ultimate hope of this program is to use the powerful apparatus of
$2$ dimensional conformal field theory in the context of quantum $4$
dimensional general relativity, for instance in an S-matrix approach
between $\scri^+$ and $\scri^-$.

\section*{Appendix 1: Central extensions of $\mathfrak{bms}_3$}

In order to get rid of the overall $i$ in (\ref{eq:3}), we redefine
the generators as $l^\prime_m=il_m$. Inequivalent central extensions
of $\mathfrak{bms}_3$ are classified by the cohomology space
$H^2(\mathfrak{bms}_3)$. More explicitly, the Chevally-Eilenberg
differential is given by
\begin{equation}
  \label{eq:4}
  \gamma=-\half C^mC^{k-m}(2m-k)\dover{}{C^k}-C^m\xi^{k-m}(2m-k) 
\dover{}{\xi^k}\,,
\end{equation}
in the space $\Lambda(C,\xi)$ of polynomials in the anticommuting
``ghost'' variables $C^m,\xi^m$. The grading is given by the
eigenvalues of the ghost number operator,
$N_{C,\xi}=C^m\dover{}{C^m}+\xi^m\dover{}{\xi^m}$, the differential
$\gamma$ being homogeneous of degree $1$ and
$H^2(\mathfrak{bms}_3)\simeq H^2(\gamma,\Lambda(C,\xi))$.
Furthermore, when counting only the ghosts $\xi^m$ associated with
supertranslations, $N_\xi=\xi^m\dover{}{\xi^m}$, the differential
$\gamma$ is homogeneous of degree $0$, so that the cohomology
decomposes into components of definite $N_\xi$ degree. The cocycle
condition then becomes
\begin{eqnarray}
\gamma(\omega^0_{m,n}C^mC^n)=0,\quad 
\gamma(\omega^1_{m,n}C^m\xi^n)=0,\quad 
\gamma(\omega^2_{m,n}\xi^m\xi^n)=0, 
\label{eq:5}
\end{eqnarray}
with $\omega^0_{m,n}=-\omega^0_{n,m}$ and
$\omega^2_{m,n}=-\omega^2_{n,m}$. The coboundary condition reads
\begin{eqnarray}
  \label{eq:6}
  \omega^0_{m,n}C^mC^n=\gamma(\eta^0_mC^m),\quad
  \omega^1_{m,n}C^m\xi^n=\gamma(\eta^1_m\xi^m). 
\end{eqnarray}

We have $\{\dover{}{C^0},\gamma\}=\cN_{C,\xi}$ with
$\cN_{C,\xi}=m(C^m\dover{}{C^m}+\xi^m\dover{}{\xi^m})$. It follows
that all cocycles of $\cN_{C,\xi}$ degree different from $0$ are
coboundaries, $\gamma \omega_N=0$, $\cN_{C,\xi}\omega_N=N\omega_N$,
$N\neq 0$ implies that
$\omega_N=\gamma(\frac{1}{N}\dover{}{C^0}\omega_N)$. Without loss of
generality we can thus assume that
$\omega^0_{m,n}C^mC^n=\omega^0_mC^mC^{-m}$ with
$\omega^0_m=-\omega^0_{-m}$ and in particular $\omega^0_0=0$;
$\omega^1_{m,n}C^m\xi^n=\omega^1_mC^m\xi^{-m}$;
$\omega^2_{m,n}\xi^m\xi^n=\omega^2_m\xi^m\xi^{-m}$ with
$\omega^2_m=-\omega^2_{-m}$ and in particular $\omega^2_0=0$. By
applying $\dover{}{C^0}$ to the coboundary condition
$\omega^0_mC^mC^{-m}=\gamma(\eta^0_mC^m)$ we find that $0=m\eta^0_m
C^m$. The coboundary condition then gives
$\omega^0_mC^mC^{-m}=\gamma(\eta^0_0 C^0)=-m \eta^0_0 C^m C^{-m}$. By
adjusting $\eta^0_0$, we can thus assume without loss of generality
that $\omega^0_1=0$ and that the coboundary condition has been
entirely used. In the same way
$\omega^1_mC^m\xi^{-m}=\gamma(\eta^1_m\xi^m)$ implies first that
$\eta^1_m=0$ for $m\neq 0$ and then that one can assume that
$\omega^1_1=0$, with no coboundary condition left.

Taking into account the anticommuting nature of the ghosts, the
cocycle conditions become explicitly,
$\omega^0_{m}(2n+m)-\omega^0_{n}(2m+n)+\omega^0_{m+n}(n-m)=0$,
$\omega^1_{m}(2n-m)+\omega^1_{n}(n-2m)+\omega^1_{m-n}(n+m)=0$,
$\omega^2_{m}(2n+m)+\omega^2_{m+n}(n-m)=0$. Putting $m=0$ in the last
relation gives $\omega^2_m=0$, for $m\neq 0$ and thus for all $m$,
putting $m=1=n$ in the second relation gives $\omega^1_0=0$, while
$m=0$ gives $\omega^1_n n=-\omega^1_{-n}n$ and thus that
$\omega^1_n=-\omega^1_{-n}$ for all $n$. Changing $m$ to $-m$ and
using this symmetry property, the cocycle conditions for $\omega^0_m$
and $\omega^1_m$ give the same constraints. Putting $m=1$, one finds
the recurrence relation
$\omega^{0,1}_{n+1}=\frac{n+2}{n-1}\omega^{0,1}_n$, which gives a
unique solution in terms of $\omega^{0,1}_2$. The result follows by
setting $c_{1,2}=\half \omega^{0,1}_2$ and checking that the
constructed solution does indeed satisfy the cocycle condition.

\section*{Appendix 2: Central extensions of $\mathfrak{bms}_4$}

For $\mathfrak{bms}_4$, the Chevally-Eilenberg differential is given
by
\begin{eqnarray}
  \label{eq:4bis}
  \gamma &=&-\half C^mC^{k-m}(2m-k)\dover{}{C^k}-\half \bar C^m\bar
  C^{k-m}(2m-k)\dover{}{\bar C^k}\nonumber
 \\ &&-C^m\xi^{k-m,n}(\frac{3m+1}{2}-k)\dover{}{\xi^{k,n}}
- \bar C^n\xi^{m,k-n}(\frac{3n+1}{2}-k)\dover{}{\xi^{m,k}}
\,,
\end{eqnarray}
in the space $\Lambda(C,\bar C,\xi)$ of polynomials in the
anticommuting ``ghost'' variables $C^m,\bar C^n,\xi^{m,n}$. The
grading is given by the eigenvalues of the ghost number operator,
$N_{C,\xi}=C^m\dover{}{C^m}+\bar C^m\dover{}{\bar
  C^m}+\xi^{m,n}\dover{}{\xi^{m,n}}$, the differential $\gamma$ being
homogeneous of degree $1$ and $H^2(\mathfrak{bms}_4)\simeq
H^2(\gamma,\Lambda(C,\bar C,\xi))$.  Furthermore, when counting only the
ghosts $\xi^m$ associated with supertranslations,
$N_\xi=\xi^{m,n}\dover{}{\xi^{m,n}}$, the differential $\gamma$ is homogeneous
of degree $0$, so that the cohomology decomposes into components of
definite $N_\xi$ degree. The cocycle condition then becomes
\begin{eqnarray}
\gamma(\omega^0_{m,n}C^mC^n+\bar\omega^0_{m,n}\bar C^m\bar
C^n+\omega^{-1}_{m,n}C^m\bar C^n)=0, \quad &&
\gamma(\omega^1_{k,mn}C^k\xi^{m,n}+\bar \omega^1_{k,mn}\bar
C^k\xi^{m,n})=0,\nonumber \\
&&\gamma(\omega^2_{mn,kl}\xi^{m,n}\xi^{k,l})=0, 
\label{eq:5bis}
\end{eqnarray}
with $\omega^0_{m,n}=-\omega^0_{n,m}, \bar \omega^0_{m,n}=-\bar \omega^0_{n,m}$ and
$\omega^2_{mn,kl}=-\omega^2_{kl,mn}$. 
The coboundary condition reads
\begin{eqnarray}
  \label{eq:6bis}
 \omega^0_{m,n}C^mC^n+\bar\omega^0_{m,n}\bar C^m\bar
C^n+\omega^{-1}_{m,n}C^m\bar C^n=\gamma(\eta^0_mC^m+\bar\eta^0_m\bar
C^m),\nonumber\\
\omega^1_{k,mn}C^k\xi^{m,n}+\bar \omega^1_{k,mn}\bar
C^k\xi^{m,n}=\gamma(\eta^1_{mn}\xi^{m,n}). 
\end{eqnarray}

We have $\{\dover{}{C^0},\gamma\}=\cN_{C,\xi}$ with
$\cN_{C,\xi}=mC^m\dover{}{C^m}+(m-\half)\xi^{m,n}\dover{}{\xi^{m,n}}$ and
also $\{\dover{}{\bar C^0},\gamma\}=\bar \cN_{\bar C,\xi}$ with
$\bar \cN_{\bar C,\xi}=n\bar C^n\dover{}{\bar
  C^n}+(n-\half)\xi^{m,n}\dover{}{\xi^{m,n}}$. 
It follows again that
all cocycles of either $\cN_{C,\xi}$ or $\bar \cN_{\bar C,\xi}$ degree different from $0$ are
coboundaries. Without loss of generality we can thus assume that
$\omega^0_{m,n}C^mC^n+\bar\omega^0_{m,n}\bar C^m\bar
C^n+\omega^{-1}_{m,n}C^m\bar C^n=\omega^0_{m}C^mC^{-m}+\bar\omega^0_{m}\bar C^m\bar
C^{-m}+\omega^{-1}_{0,0}C^0\bar C^0$ with 
$\omega^0_m=-\omega^0_{-m}, \bar \omega^0_m=-\bar \omega^0_{-m}$ and
in particular $\omega^0_0=0=\bar \omega^0_0$; none of monomials with
one $\xi^{m,n}$ and either on $C^k$ or one $\bar C^k$ can be of degree $0$, so 
$\omega^1_{k,mn}=0=\bar \omega^1_{k,mn}$;
$\omega^2_{mn,kl}\xi^{m,n}\xi^{k,l}=\omega^2_{m,n}\xi^{m,n}\xi^{-m+1,-n+1}$ with 
$\omega^2_{m,n}=-\omega^2_{-m+1,-n+1}$. Both the cocycle and the
coboundary condition for $\omega^0_{m}C^mC^{-m}+\bar\omega^0_{m}\bar C^m\bar
C^{-m}+\omega^{-1}_{0,0}C^0\bar C^0$ split. For
$\omega^{-1}_{0,0}C^0\bar C^0$ there is no coboundary condition, while 
the cocycle condition implies $\omega^{-1}_{0,0}=0$. The rest of the
analysis proceeds as in the previous subsection, separately for
$\omega^0_{m}C^mC^{-m}$ and $\bar\omega^0_{m}\bar C^m\bar
C^{-m}$, with the standard central extension for $[l_m,l_{-m}]$ and
$[\bar l_m,\bar l_{-m}]$.

We still have to analyze
$\gamma(\omega^2_{m,n}\xi^{m,n}\xi^{-m+1,-n+1})=0$. This condition
gives
$\omega^2_{m,n}(\frac{3l-1}{2}+m)+\omega^2_{l+m,n}(\frac{l+1}{2}-m)=0$
and also
$\omega^2_{m,n}(\frac{3l-1}{2}+n)+\omega^2_{m,l+n}(\frac{l+1}{2}-n)=0$. 
Putting $m=0$ in the first relation gives
$\omega^2_{0,n}(\frac{3l-1}{2})+\omega^2_{l,n}(\frac{l+1}{2})=0$. Putting
$l=-1$ then implies $\omega^2_{0,n}=0$ and then also
$\omega^2_{l,n}=0$ for $l\neq -1$. But
$\omega^2_{-1,n}=-\omega^2_{2,-n+1}=0$ which shows that
$\omega^2_{m,n}=0$ for all $m,n$ and concludes the proof.

\section*{Acknowledgments}

The authors are grateful to G.~Comp\`ere and P.-H.~Lambert for
collaborations on these topics and to M. Henneaux for a useful
discussion. This work is supported in part by the Fund for Scientific
Research-FNRS (Belgium), by the Belgian Federal Science Policy Office
through the Interuniversity Attraction Pole P6/11, by IISN-Belgium, by
``Communaut\'e fran\c caise de Belgique - Actions de Recherche
Concert\'ees'' and by Fondecyt Projects No.~1085322 and
No.~1090753. The participation in the workshop was partially financed
by the Quantum Geometry and Quantum Gravity Research Networking
Programme of the European Science Foundation.

%\bibliography{/Users/gbarnich/Documents/Literature/Bibliography/master2}

\begin{thebibliography}{10}

\bibitem{Barnich:2006avcorr}
G.~Barnich and G.~Comp{\`e}re, ``Classical central extension for asymptotic
  symmetries at null infinity in three spacetime dimensions,'' {\em Class.
  Quant. Grav.} {\bf 24} (2007) F15,
  \href{http://www.arXiv.org/abs/gr-qc/0610130}{{\tt gr-qc/0610130}}.
Corrigendum: ibid 24 (2007) 3139.
%%CITATION = GR-QC/0610130;%%.

\bibitem{Hollands:2003ie}
S.~Hollands and A.~Ishibashi, ``{Asymptotic flatness and Bondi energy in higher
  dimensional gravity},'' {\em J. Math. Phys.} {\bf 46} (2005) 022503,
\href{http://www.arXiv.org/abs/gr-qc/0304054}{{\tt gr-qc/0304054}}.
%%CITATION = GR-QC/0304054;%%.

\bibitem{Tanabe:2009va}
K.~Tanabe, N.~Tanahashi, and T.~Shiromizu, ``{On asymptotic structure at null
  infinity in five dimensions},''
\href{http://www.arXiv.org/abs/0909.0426}{{\tt 0909.0426}}.
%%CITATION = 0909.0426;%%.

\bibitem{Brown:1986nw}
J.~D. Brown and M.~Henneaux, ``Central charges in the canonical realization of
  asymptotic symmetries: An example from three-dimensional gravity,'' {\em
  Commun. Math. Phys.} {\bf 104} (1986) 207.

\bibitem{fefferman:1985}
C.~Fefferman and C.~Graham, {\em {Elie Cartan et les Math\'ematiques
  d'aujourd'hui}}, ch.~{Conformal Invariants}, pp.~95--116.
\newblock Ast{\'e}risque, 1985.

\bibitem{graham:1991}
C.~Graham and J.~Lee, ``{Einstein metrics with prescribed conformal infinity on
  the ball},'' {\em Adv. Math.} {\bf 87} (1991) 186--225.

\bibitem{Barnich:2009se}
G.~Barnich and C.~Troessaert, ``{Symmetries of asymptotically flat 4
  dimensional spacetimes at null infinity revisited},'' {\em Phys. Rev. Lett.}
  {\bf 105} (2010) 111103,
\href{http://www.arXiv.org/abs/0909.2617}{{\tt 0909.2617}}.
%%CITATION = 0909.2617;%%.

\bibitem{Barnich:2010eb}
G.~Barnich and C.~Troessaert, ``{Aspects of the BMS/CFT correspondence},'' {\em
  JHEP} {\bf 05} (2010) 062,
\href{http://www.arXiv.org/abs/1001.1541}{{\tt 1001.1541}}.
%%CITATION = 1001.1541;%%.

\bibitem{Bondi:1962px}
H.~Bondi, M.~G. van~der Burg, and A.~W. Metzner, ``Gravitational waves in
  general relativity. 7. {W}aves from axisymmetric isolated systems,'' {\em
  Proc.\ Roy.\ Soc.\ Lond. A} {\bf 269} (1962)
21.
%%CITATION = PRSLA,A269,21;%%.

\bibitem{Sachs:1962wk}
R.~K. Sachs, ``Gravitational waves in general relativity. 8. {W}aves in
  asymptotically flat space-times,'' {\em Proc.\ Roy.\ Soc.\ Lond.\ A} {\bf
  270} (1962)
103.
%%CITATION = PRSLA,A270,103;%%.

\bibitem{Sachs2}
R.~K. Sachs, ``Asymptotic symmetries in gravitational theories,'' {\em Phys.\
  Rev.} {\bf 128} (1962) 2851--2864.

\bibitem{PhysRevLett.10.66}
R.~Penrose, ``Asymptotic properties of fields and space-times,'' {\em Phys.
  Rev. Lett.} {\bf 10} (1963), no.~2, 66--68.

\bibitem{penrose:1964}
R.~Penrose, ``Conformal treatment of infinity,'' in {\em Relativity, groups and
  topology: Les Houches 1963}, B.~D. C.~DeWitt, ed., pp.~563--584.
\newblock Gordon and Breach, 1964.

\bibitem{Penrose:1974}
R.~Penrose, ``{Relativistic Symmetry Groups},'' in {\em Group theory in
  non-linear problems}, A.~O. Barut, ed., pp.~1--58.
\newblock Reidel Publishing Company, Dodrecht, Holland, 1974.

\bibitem{newman:891}
E.~T. Newman and T.~W.~J. Unti, ``Behavior of asymptotically flat empty
  spaces,'' {\em Journal of Mathematical Physics} {\bf 3} (1962), no.~5,
  891--901.

\bibitem{newman:566}
E.~Newman and R.~Penrose, ``An approach to gravitational radiation by a method
  of spin coefficients,'' {\em Journal of Mathematical Physics} {\bf 3} (1962),
  no.~3, 566--578.

\bibitem{newman:1980xx}
E.~P. Newman and K.~P. Tod, ``{Asymptotically Flat Space-times},'' in {\em
  {General Relativity and Gravitation. 100 Years after the Birth of Albert
  Einstein. Volume 2}}, {Plenum Press}, ed., pp.~1--36.
\newblock 1980.

\bibitem{barnich:2011ty}
G.~Barnich and P.-H. Lambert, ``{A note on the Newman-Unti group},''
\href{http://www.arXiv.org/abs/1102.0589}{{\tt 1102.0589}}.
%%CITATION = 1102.0589;%%.

\bibitem{Barnich:2010xq}
G.~Barnich, ``{A note on gauge systems from the point of view of Lie
  algebroids},'' {\em AIP Conf. Proc.} {\bf 1307} (2010) 7--18,
\href{http://www.arXiv.org/abs/1010.0899}{{\tt 1010.0899}}.
%%CITATION = 1010.0899;%%.

\bibitem{Ashtekar:1996cm}
A.~Ashtekar, J.~Bicak, and B.~G. Schmidt, ``{Behavior of Einstein-Rosen waves
  at null infinity},'' {\em Phys. Rev.} {\bf D55} (1997) 687--694,
\href{http://www.arXiv.org/abs/gr-qc/9608041}{{\tt gr-qc/9608041}}.
%%CITATION = GR-QC/9608041;%%.

\bibitem{Ashtekar:1996cd}
A.~Ashtekar, J.~Bicak, and B.~G. Schmidt, ``Asymptotic structure of symmetry
  reduced general relativity,'' {\em Phys. Rev.} {\bf D55} (1997) 669--686,
\href{http://www.arXiv.org/abs/gr-qc/9608042}{{\tt gr-qc/9608042}}.
%%CITATION = GR-QC 9608042;%%.

\bibitem{Fuks:1986}
D.~Fuks, {\em {Cohomology of infinite-dimensional Lie algebras}}.
\newblock Consultants Bureau, New York, 1986.

\bibitem{Henneauxstrings}
L.~Brink and M.~Henneaux, {\em Principles of string theory}, ch.~II: {L}ectures
  on string theory, with emphasis on Hamiltonian and BRST methods.
\newblock Plenum, New York, 1988.

\bibitem{deAzcarraga:1989dm}
{Jose A. de Azc\'arraga and Jos\'e M. Izquierdo}, {\em Lie groups, Lie
  algebras, cohomolgy, and some applications in physics}.
\newblock Cambridge University Press, 1995.

\bibitem{PhysRevLett.105.171601}
A.~Bagchi, ``Correspondence between asymptotically flat spacetimes and
  nonrelativistic conformal field theories,'' {\em Phys. Rev. Lett.} {\bf 105}
  (Oct, 2010) 171601.

\bibitem{Bagchi:2009pe}
A.~Bagchi, R.~Gopakumar, I.~Mandal, and A.~Miwa, ``{GCA in 2d},'' {\em JHEP}
  {\bf 08} (2010) 004,
\href{http://www.arXiv.org/abs/0912.1090}{{\tt 0912.1090}}.
%%CITATION = 0912.1090;%%.

\bibitem{winicour:1980aa}
J.~Winicour, {\em {General Relativity and Gravitation. 100 Years after the
  Birth of Albert Einstein.}}, vol.~2, ch.~{3. Angular Momentum in General
  Relativity}, pp.~71--93.
\newblock New York: Plenum, 1980.

\end{thebibliography}

\def\cprime{$'$}
\providecommand{\href}[2]{#2}\begingroup\raggedright\endgroup

\end{document}